\documentstyle[prl,aps,epsfig]{revtex}
\draft
\begin{document}
\twocolumn[
\hsize\textwidth\columnwidth\hsize\csname@twocolumnfalse\endcsname

\hskip 9cm\vskip -1cm ETH-TH/99-19

\title{The underscreened Kondo effect in ladder systems}
\author{Karyn Le Hur}
\address{Theoretische Physik, ETH-H\"onggerberg, CH-8093 Z\"urich, Switzerland}
 \maketitle

\begin{abstract}

We introduce an unfamiliar Kondo ladder problem by coupling the well-known
Takhtajan-Babujan S=1 chain to a half-filled one-dimensional electron
gas, and we solve it using quantum field theory techniques.
Based on the so-called underscreened
Kondo effect, we find an insulating spin liquid with both
deconfined spinons and optical magnons. Important consequences are discussed.
\end{abstract}
\pacs{PACS numbers: 71.10.Pm, 71.27.+a, 75.10.Jm} 
\twocolumn
\vskip.5pc ]
\narrowtext

The problem of the co-existence between conduction electrons and 
an array of local
magnetic moments remains an eminent unsolved problem in condensed
matter physics, in link with the well-known Kondo
effect. This poses a real challenge to the theorist.
As a first step, the one-dimensional
(1D) model of a Kondo chain with an M-fold 
degenerate band of conduction electrons interacting with a spin S
Heisenberg chain has been solved in a semi-classical limit\cite{Tsvelik}. At 
low energy, the system is insulating and the spin
properties can be described by an 0(3) nonlinear sigma model (NL$\sigma$M)
with a topological term $\theta=\pi(2S-M)$. For $\left|M-2S\right|=$ (even), 
single electron excitations at low energies are massive spin polarons. For
$\left|M-2S\right|=$ (odd), the NL$\sigma$M in contrast becomes 
critical and belongs to the 
universality class of the isotropic S=1/2 Heisenberg chain. 
In this Letter, we
investigate the interesting case S=1 and M=1 in a so-called
`ladder' picture\cite{Maurice}.
As we shall see below, the 
characterization of the spin array as an S=1
quantum spin chain with prevalent gapless triplet excitations 
should allow to
push forward our understanding of the underscreened Kondo effect
\cite{Nozieres} in the Kondo lattice model (KLM). 
For experimental applicability, this
model appears to be particularly appropriate to describe 
heavy-fermion materials with Uranium compounds such as
$UPt_3$ (as pointed
out in ref.\cite{KLH1}, where we have firstly solved the problem of two 
localized S=1 spins embedded in a host metal). 

Let us start with the integrable Takhtajan-Babujan S=1 chain on a lattice,
described by:
${\cal H}=J_{\parallel}\sum_j ({\bf S}_j{\bf S}_{j+1})-
\beta({\bf S}_j{\bf S}_{j+1})^2$
where $\beta=1$\cite{TB}. ${\bf S}_j$ describe local moments.
The unusual term quartic in spin can be generated, for 
example, by phonons\cite{Nagaev}. Solving Bethe Ansatz
equations, the only elementary excitation is known to be a doublet
of gapless 
spin-1/2 spin waves, with total spin 0 or 1\cite{TB}. 
The approach discussed in 
this Letter is based on field theory techniques
like the non-Abelian\cite{aff} and  Abelian\cite{Haldane} 
bosonizations. Thus, it is convenient
to use a (simple) {\it continuum} description of such an integrable model.
Interestingly, the specific
heat behaves as: $C_v/L=2S T/(1+S)+{\cal O}(T^3)$, where S=1\cite{TB}. 
In the sense
of critical theories, such a model could be parametrized by a conformal
anomaly (which is the central charge in the Virasoro algebra):
$C=\frac{3S}{S+1}=3/2$\cite{aff2}. This fact, together with 
approximate mappings, leads to
the conjecture that the criticality of this model is governed by operators
satisfying a 
critical Wess-Zumino-Witten (WZW) model on the SU(2) group at the level 
$k=2S=2$\cite{AH}. Solving numerically the Bethe-Ansatz equations
the authors of ref.\cite{Alcaraz} checked
the main conjectures made in refs.\cite{aff2,AH}. Thus, for $\beta=1$,
the effective action for
the low-energy excitations is given by the WZW model
on the SU(2) group at the level $k=2$.
The following general statement is valid: the
Kac-Moody algebra for the group SU(N) at the level k=N can be represented
by the currents of Majorana fermions $\xi^a(a=1,...,N^2-1)$ from
the adjoint representation of this group. Thus, the $SU(2)_{k=2}$ currents 
can be explicitly
rewritten as:
$\hbox{I}^a({\bar{\hbox{I}}}^a)=-\frac{i}{2}\epsilon^{abc}\xi_{L(R)}^b
\xi_{L(R)}^c$.
The Hamiltonian is in turn equivalent to the model of 
three massless Majorana fermions $\xi^a(a=1,2,3)$\cite{alexei}.
In the continuum limit, the spin-1 chain has {\it only} zero
sound triplet excitations (or {\it magnons}, like in conventional magnets). 
The staggered
magnetization ${\bf N}$ can be simply 
expressed as ${\bf N}(x)\simeq 
\hbox{Tr}\{(h+h^{\dag})\sigma\}$, where $h\in SU(2)_{k=2}$ has dimension 3/8
\cite{AH,alexei}. 

The spectrum of {\it free} conduction electrons yields a separation of 
spin and 
charge. The 
spin sector is described
by an SU(2) WZW model with k=1\cite{aff}. Here, the two 
fundamental fields
$g$, $g^{\dag}$ are {\it spinon} pairs which can be
viewed as free
fields except from a purely statistical (in that case: semionic)
interaction. The electronic spin density is usually described as: ${\bf
  S}_c(x)={\bf J}_c(x)+e^{2ik_Fx}{\bf N}_c(x)$, where ${\bf J}_c={\bf
  J}+\bar{{\bf J}}$ and ${\bf N}_c\simeq\hbox{Tr}\{{\mathbf\sigma}
(g+g^{\dag})\}\cos{\sqrt{2\pi}\Phi_c^c}$ are,
respectively, the smooth and staggered parts of the magnetization. 
${\bf N}_c$ has the dimension 1 and ${\bf J}_c$ can be 
explicitly written
as a function of $g$ and $g^{\dag}$\cite{aff}.
The charge sector is 
described
in terms of a U(1) scalar field $\Phi_c^c$ leading to a Conformal Field Theory
(CFT) with C=1 for holons as well. Zero sound
charge modes propagate with the velocity $v_{\rho}$ and are 
ruled by the Luttinger parameter $K_{\rho}=v_F/v_{\rho}$. For a 
1D free electron gas $K_{\rho}=1$, and  $v_F=2t\sin(k_Fa)$ is the velocity for 
charge and spin degrees of freedom. The Kondo interaction takes the form:
\begin{eqnarray}
\label{zero}
{\cal H}_{int}&=&\lambda_2({\bf J}+\bar{{\bf J}})({\bf I}+\bar{{\bf I}})
\\ \nonumber
&+&\alpha\lambda_3e^{i\delta x}\ \hbox{Tr}(g^{\dag}\sigma)\cos\sqrt{2\pi}
{\Phi}_c^c{\bf N}+h.c.
\end{eqnarray}
where $\lambda_{2,3}\propto J_K$, and $\alpha$ is a non-universal constant.
We have defined $\delta=2k_F-\pi$. Here, we consider
situations of the half-filled case, with $\delta\rightarrow 0$. The bare
conditions are fixed to: $J_K\ll(J_{\parallel},t)$. The coupling between the 
short-range staggering fluctuations becomes there prominent. A small term 
${\lambda_3}$ modifies the universal dc conductivity of
the electron gas $\sigma_o=2e^2K_{\rho}L/h$, obtained by applying a
static field over a finite part of the pure sample. In the presence of
magnetic impurities, the dc conductivity is rescaled as $\sigma(T)\propto
\{T\lambda_3(T)^2\}^{-1}$\cite{KLH2}. We assume that the phase coherence is 
lost after each magnetic collision. Then, the diffusion by the array of 
magnetic
impurities with S=1 provides an anomalous power-law, $\sigma(T)\propto
T^{1/4}$. Note that the smallest amount of Kondo `disorder' will destroy the
so-called Umklapp term which gives $\sigma(T)\sim T^{-1}$ (when $K_{\rho}=1$) 
by using a memory-function approximation\cite{Giam1}. Clearly, the effects of 
scattering from the array of magnetic impurities become coherent at lower 
energy which
typically give rise to a Kondo localization. An estimate for the localization
length may be simply obtained from one-loop perturbation theory; it gives:
${\cal L}_{mag}=(1/J_K)^{8/5}$. 

The dc conductivity for this state can be easily computed using the Kubo
formula. The resistivity follows $\rho(T)\propto e^{\Delta/k_B T}$ when
$\hbox{k}_B T\ll \Delta$ leading to a perfect insulating state. When
$\delta=0$, the 
mass gap (which can be identified either as the charge gap or
the spin gap since $\lambda_3$ couples spin and charge sectors) 
obeys $\Delta={\cal L}_{mag}^{-1}\equiv {J_K}^{8/5}$. 
It is larger than in the usual 1D
KLM (with S=1/2) for the same bare conditions\cite{KLH2}.
 The only possibility to obtain a non-zero conductivity
is through thermal fluctuations which can activate
charge excitations. When
$T\rightarrow 0$, correlations of $\cos\sqrt{2\pi}{\Phi}_c^c$ show a 
long-range order. Summing the
spin part and refermionizing the theory, 
the diagonalization of the charge sector is an easy task
even for $\delta\neq 0$. We find a gas of spinless fermions ruled by the 
dispersion relation: $\epsilon(k)=\pm\sqrt{v_F^2 k^2+\Delta^2}- 
v_F\delta/2$.  
It rescales the mass gap as $\Delta(\delta)=\Delta-v_F\delta/2$ and the
insulating state can be stabilized as long as
$\Delta\gg v_F\delta/2$.

Now, let us investigate the spin properties of such an
insulating Kondo system (for $\delta=0$) and explain the
crucial idea of the paper.
The physics behind these
calculations is the formation of bound states between conduction electrons 
and localized spins, leading to a semi-conducting phase. Of course, the
occurrence of Kondo bound states
can be achieved only due to the splitting of each S=1 localized spin onto
two doublet modes. Note that although powerful
techniques have been used to solve the S=1 impurity Kondo model, it is
difficult to have a
`simple' mathematical formulation of the splitting phenomenon
(see ref.\cite{KLH1} and references therein).
In the present lattice model, we express it simply as 
follows. For $T\simeq \Delta$ (typically the spin gap) 
we use the decomposition,
\begin{equation}
\label{star}
{\bf N}\Delta^{1/8}=({\bf N}_1+{\bf N}_2)
\end{equation}
with ${\bf N}_i=\hbox{Tr}(h_i\sigma)$ and $h_i\in SU(2)_{k=1}$. 
Through the factor $\Delta^{1/8}$, we 
consider that the operators $h_i$ (with k=1)
have the 
scaling dimension
1/2 and not 3/8. Only `half' of each S=1 spin can form a bound state with a
conduction electron. Then, one doublet is assumed to participate 
in the Kondo
process whereas the other would remain massless. 
This phenomenon can be simply described through the important
formula: 
\begin{equation}
\label{cinq}
\frac{\bf N}{2}\hbox{Tr}(g\sigma)=\Delta^{-1/8}\ \hbox{Tr}(h_1^{\dag}\sigma)
\hbox{Tr}(g\sigma)
\end{equation}
First, the SU(2) field $h_2$ remains free (except statistical
interactions), leading to a critical C=1 CFT (i.e. a Heisenberg model)
with deconfined spinons in
the ground state. Then, despite of the insulating state the optical 
conductivity would reveal only a pseudogap at $k=\pi$ through
spin instantons. Following
the arguments of Tsvelik\cite{Tsvelik}, one can define an anomalous 
fermionic 
current ${\bf
J}_A$ which is proportional to the topological charge density of the
staggered magnetization ${\bf N}_2$. Thus, ${\bf J}_A$ varies linearly
with $\frac{1}{\pi}\epsilon^{\mu\nu}{\bf N}_2(\partial_{\mu}{\bf
N}_2\times\partial_{\nu}{\bf N}_2)$ which has the scaling
dimension 5/2. In the
$(q,w)$ space, it gives a purely dynamical optical conductivity
$\sigma_{opt}(w,k=\pi-q)\propto\frac{1}{w}(w^2-v_F^2q^2)^{3/2}\sim w^2$. 
Second, deconfinement of
spinons does not imply that conventional magnons fail to be
quasiparticles in such a spin liquid. The low-energy physics also involves
optical magnons. 
It is convenient to define:
$\eta=\lambda_3\langle\cos\sqrt{2\pi}{\Phi}_c^c(x)\rangle\propto
\Delta^{9/8}$. During the spin-1 splitting mechanism, the
Kondo exchange is rescaled; we change, $2\eta\rightarrow \eta$. Then, by 
using Eq.(\ref{cinq}), the dimensional equality 
$\eta_0=\eta\Delta^{-1/8}\simeq\Delta$ and omitting
high-energy singlet magnon modes (see later), we find a massive part 
described by:
\begin{equation}
S_{U}=\int d^2x\ \frac{1}{2\gamma}\hbox{Tr}(\nabla U\nabla U^{\dag})-\eta_0 :
\hbox{Tr}(U+U^{\dag}):
\end{equation}
where,
${\gamma}^{-1}\propto\sqrt{v_F/J_{\left|\right|}a}+
\sqrt{J_{\left|\right|}a/v_F}$,\ 
$x_0=\tau\sqrt{v_FJ_{\left|\right|}a}$,\ $x_1=x$ and $U(x)=gh_1^{\dag}(x)$ 
has
SU(2) symmetry.
\vskip 0.1cm
Excitations of the O(3) NL$\sigma$M are S=1 triplets (magnons).
At low temperatures, $\hbox{Tr} U(x)=\hbox{constant}$ and then magnons become
purely massive.  Since by construction $\hbox{Tr}U$ has the scaling dimension
1, the spin gap (of course) is of order
of $\eta_0\simeq\Delta$. The action $S_U$ can be rewritten as a triplet of 
Majorana 
fields $\chi^a(a=1,2,3)$ with the same mass
$m_t=\Delta$. They mainly contribute to a coherent $\delta$-peak to the 
dynamical
susceptibility $\chi''(q=\pi -k>0,\omega)\sim
m_t{(q^2+m_t^2)}^{-1/2}\delta(\omega-\sqrt{q^2+m^2_t})$ near $k=\pi$ and
$\omega=m_t$.  On the other hand, the spectral
density of the staggered magnetization $(-1)^x {\bf N}_2(x)$ with ${\bf
N}_2\sim \hbox{Tr}(h_2\sigma)$ shows a purely incoherent background. 
Schulz\cite{Heinz1}
explained that it leads to: $\chi''(q>0,\omega)\simeq\sin (2\pi c) 
T^{4c-2}\Im
m[\rho(\frac{w-q}{4\pi T})\rho(\frac{w+q}{4\pi T})]$ with
$\rho(x)=\frac{\Gamma(c-ix)}{\Gamma(1-c-ix)}$ and c=1/4 for the isotropic
Heisenberg chain. Such a contribution diverges when $\omega\rightarrow q$ as
$(w-q)^{-1}$ so that the $\delta$-peak due to optical magnons remains
distinguishable at $w=\sqrt{q^2+m^2_t}$. 

During the Kondo localization, the current ${\bf \hbox{I}}$ is also
decomposed into two independent parts $({\bf J}_1,{\bf J}_2)$ where 
${\bf J}_i$ are SU(2) currents at the level k=1 (with C=1
each). Some degrees of freedom
should account for the missing C=1/2. Those are associated with the discrete 
$({\cal Z}_2)$ nonmagnetic interchange symmetry
$1\leftrightarrow 2$ in
the definition of ${\bf I}$.
 The marginal term can be rewritten: $\lambda_2({\bf J}_c+\bar{\bf
J}_c)({\bf J}_1+\bar{\bf J}_1)$. We have the correspondence 
$({\bf J}_1\oplus{\bf J}_c)=-\frac{i}{2}
\epsilon^{abc}\chi_{L}^b
\chi_{L}^c$. Then, by using the Dyson equation for the fields $\chi_i$, we 
find that the mass $m_t$ is slightly renormalized: $\Delta(1
+\frac{J_Ka}{2\pi v_F}\ln
\frac{\Lambda}{\Delta})$ and $\Lambda\sim t$. Next, we neglect
these logarithmic corrections. 
For $J_{\parallel}a\neq v_F$, free magnon excitations can survive for
$T<\Delta$. We can integrate out excitations which correspond to
configurations where $\hbox{Tr} U\not=0$ acquires a gap. It becomes an O(3)
NL$\sigma$M where the topological term has no contribution. Finally, it 
provides a rescaled mass gap
$m_t\sim\Delta\exp(-2\pi\tilde{\gamma}^{-1})$, where
$\tilde{\gamma}^{-1}\propto \left|J_{\parallel}-t\right|$. 
Here, the spin correlation
functions of the massive part are proportional to the Mac-Donald function 
$K_0(m_t {\bf r})$ with ${\bf r}=(x_0,x_1)$ and
$K_0(x)=\sqrt\frac{\pi}{2}x^{-1/2}e^{-x}(1-1/8x+{\cal
O}(x^{-2}))$\cite{Smirnov}. Massive polarons should occur in the 
interval between $m_t$ and $\Delta$ due to interactions of electrons and 
kinks in ${\bf N}_1$\cite{Tsvelik}. 

The conclusion that the fixed point is
the k=1 WZW model is consistent with Zamalodchikov's c-theorem\cite{Zamo}. 
This theorem
states that if a field theory flows between two conformally invariant fixed
points, then the value of the conformal anomaly parameter C must be smaller 
at
the Infra Red stable fixed point. Here, the
unstable fixed point is ruled by $C=7/2$; the stable fixed point
has $C=1$. Deconfined spinons remain as a result of the non-Abelian chiral
symmetry which is not broken. Note that the spinon velocity $v\sim 
J_{\parallel}a$ 
is equal to that in the single S=1/2 spin chain problem\cite{Mathias}.
Through the transform 
$\xi^i\rightarrow \chi^i$,
magnons become optical. For $T\ll \Delta$, only the gapless
modes (spinons) contribute to the specific heat $C_v=\pi T/3 v$. 
Now, we choose the particular point: $J_{\parallel}a=v_F$. Using the Abelian 
bosonization we proceed to calculate quantities of interest like
the spin-spin correlations, the NMR relaxation rate ${\cal T}_1$ and the 
magnetization by varying the external magnetic field. 

First, notice that an 
alternative
picture of the integrable S=1 chain with $\beta=1$ is realized in the
context of spin ladders, when adding a strong four-spin interaction $V\sum_j
{\bf S}_1(j){\bf S}_1(j+1){\bf S}_2(j){\bf S}_2(j+1)$ equal to the direct
transverse exchange $J_{\perp}<0$\cite{Ners}. Second, using the common 
bosonization scheme 
for the two-leg ladder (see, for instance, Parts II and III of 
ref.\cite{Shelton}), 
we
can rewrite the magnetization ${\bf N}$ of the S=1 chain in terms of two
 scalar fields 
$\Phi_e$ and $\Phi_o$, and their duals $\Theta_e$
and $\Theta_o$, as follows: 
\begin{eqnarray}
\hbox{N}_x&=&\cos\sqrt{\pi}\Theta_e\cos\sqrt{\pi}\Theta_o\\ \nonumber
\hbox{N}_y&=&\sin\sqrt{\pi}\Theta_e\cos\sqrt{\pi}\Theta_o \\ \nonumber
\hbox{N}_z&=&\sin\sqrt{\pi}\Phi_e\cos\sqrt{\pi}\Phi_o
\end{eqnarray}
For $J_{\perp}=V$, note that $\Phi_e$ and $\Theta_e$ are
completely {\it free}, whereas the Hamiltonian of $\Phi_o$ describes a single 
free Majorana field. It leads to a CFT with a total central charge C=3/2. The 
operators $\exp i\sqrt{\pi}\Phi_e(\Theta_e)$ and 
$\cos\sqrt{\pi}
\Phi_o(\Theta_o)$ acquire the scaling dimensions 1/4 and 1/8,
respectively. Now, we
use the representation
\begin{equation}
{\bf N}_c=(\cos\sqrt{2\pi}\Theta_c,\ \sin\sqrt{2\pi}\Theta_c,\ 
\sin\sqrt{2\pi}\Phi_c)
\end{equation}
Then, it is not difficult to show that the splitting of each S=1 local spin 
can be 
still understood through the formula (\ref{star}) where ${\bf N}_{i=1,2}$ has
the same form as ${\bf N}_c$, but replacing $\Phi_c\rightarrow \Phi_i$ and 
$\Theta_c\rightarrow \Theta_i$. We have the constraints,
$\Phi_{1,2}=1/\sqrt{2}(\Phi_e\pm\Phi_o)$. By using the Abelian 
bosonization, Eq.(\ref{cinq}) can be understood as
follows: only one bosonic field, namely $\Phi_{1}$, can hybridize with 
$\Phi_{c}$ and similarly for the dual fields. The term 
$\lambda_3$ can be simply written as:
\begin{eqnarray}
{\cal H}_{int}&=&-\eta_1:\cos\sqrt{4\pi}\Phi_+(x):+\eta_2:\cos\sqrt{4\pi}
\Phi_-(x):\\
\nonumber
&+& 2\eta_3:\cos\sqrt{4\pi}\Theta_-(x):
\end{eqnarray}
with the notations: $\eta_i\propto \Delta/\pi^2a$ and
$\Phi_{+,-}=1/\sqrt{2}(\Phi_1\pm\Phi_c)$. The physics depicted by 
${\cal H}_{int}$
becomes similar to that of the two-leg ladder system, with $V=0$. 
\vskip 0.1cm
Results obtained
in ref.\cite{Shelton} are then applicable to that case and we present them 
briefly 
for completness. The Abelian bosonization method gives an {\it exact} massive 
spectrum
which is composed of a triplet of Majorana fields $\chi^a(a=1,2,3)$
with the same mass $m_t=\Delta$, and a singlet $\chi^o$ branch 
shifted to the high energy $m_s=-3\Delta$. The total ${\bf N}^{(+)}$ and 
relative ${\bf
N}^{(-)}$ staggered magnetization of the {\it massive} part are given by,
\begin{eqnarray}
{\bf N}^{(+)}&=&\{\sigma_1\mu_2\sigma_3\sigma,\ \mu_1\sigma_2\sigma_3\sigma,
\ \sigma_1\sigma_2\mu_3\sigma\}\\ \nonumber
{\bf N}^{(-)}&=&\{\mu_1\sigma_2\mu_3\mu,\ \sigma_1\mu_2\mu_3\mu,\
\mu_1\mu_2\sigma_3\mu\}
\end{eqnarray}
where $\sigma_i,\mu_i$ are order and disorder parameters of three Ising 
models
corresponding to the massive triplet of the Majorana fields $(\chi^a)$ while
$\sigma,\mu$ refer to the singlet branch $(\chi^o)$. Masses determine
deviations from criticality such as $m_{t(s)}\equiv(T-T_c)/T_c>\hskip
-0.1cm(<)\hskip 0.1cm 0$. It corresponds to three disordered phases with
$\langle\mu_j\rangle\neq 0$ and $\sigma_j=0$, and one ordered phase with
$\langle\mu\rangle\neq 0$ and $\sigma=0$. We used the so-called
Kramers-Wannier duality $t\rightarrow -t$, $\sigma\leftrightarrow\mu$ for 
$(\sigma$,$\mu)$ corresponding to a
negative mass. Using the asymptotic behaviors
of the different 2-point Ising correlation functions\cite{Shelton} and the 
expression
of ${\bf N}_2$, one obtains:
\begin{eqnarray}
\langle {\bf N}^{(-)}({\bf r}){\bf N}^{(-)}({\bf
0})\rangle&\propto&\frac{1}{r^{5/2}}e^{-m_t r}+{\cal O}(e^{-3m_tr})\\ 
\nonumber
\langle {\bf N}^{(+)}({\bf r}){\bf N}^{(+)}({\bf 0})\rangle&\simeq&\frac{1}
{{r}^{3/2}}e^{-2(m_t+m_s)r}\\ \nonumber
\langle {\bf N}_{2}({\bf r}){\bf N}_{2}({\bf 0})\rangle &\simeq&\frac{1}{
r}
\end{eqnarray}
For $J_K>0$, the S=1 magnetization is essentially
determined by the {\it relative} staggered part ${\bf N}^{(-)}={\bf N}_1-
{\bf N}_c$. The singlet branch gives only high-energy corrections so that
the non-Abelian action $S_U$ becomes exact at low temperatures. The fields  
$\Phi_2$ and $\Theta_2$ remain free. Thus, ${\bf N}_2$ yields
the typical correlation functions of the S=1/2 isotropic Heisenberg
 chain. A related quantity of 
direct experimental relevance is the NMR relaxation rate 
${\cal T}_1$. It is easy
to see that the staggered susceptibilities dominate in ${{\cal T}_1}^{-1}$. 
Precisely, we obtain: $\chi(2k_F)=1/T+T^{-1/2}\exp -\Delta/T$. For a 
system of 
non-interacting spins $\chi(2k_F)=({\cal T}_1 T)^{-1}$ is a 
constant. 
Here, the two-component underlying magnetic order
changes the effective field seen by the nuclei
and, for instance, ${\cal T}_1^{-1}$ goes to a nonzero value as 
$T\rightarrow 0$.

Pleasantly, as in three-coupled spin chain systems\cite{Andreas}, there 
is a phenomenon which is
strikingly analogous to the quantum Hall effect. The 
infinite degeneracy of a Heisenberg chain is lifted for a
strong magnetic field. The system
shows a plateau in the magnetization curve at magnetization {\it per site} 
m=1/2 and remains insulating even when a finite
magnetization is induced by the external field. Thus, we obtain the 
same physics as for a 
partially magnetized $S_{tot}=3/2$
chain\cite{Oshi}. Regarding each $S_{tot}=3/2$ operator as
a symmetrized product of three S=1/2's, one S=1/2 is polarized by
the the applied field at each site while the other two form a
valence-bond-state having a nonlocal topological order associated with
the breakdown of a hidden discrete ${\cal Z}_2\times{\cal Z}_2$
symmetry. Here,
we have a
two-chain realization of such a Haldane state and the `string' 
 order parameter is defined as 
$\hbox{lim}_{\left|x-y\right|\rightarrow +\infty}\langle
{\cal O}_{str}(x,y)\rangle \simeq {\langle \mu_1\rangle}^2{\langle 
\mu_2\rangle}^2\neq 0$. But, we expect the vanishing of the Haldane
 gap for some value $H=H_{c1}$ 
of the magnetic field. We closely follow the notations of ref.
\cite{Chitra}. We can see easily that $H$ affects only the symmetric field
$\Phi_{+}$ as, ${\cal H}_+
\rightarrow {\cal H}_+ +\sqrt{2}H\partial_x\Phi_+$. As long as $H<H_{c1}$
 the result is a 
Sine-Gordon model with a gap which is rescaled
by the magnetic field. It can be 
viewed as a completely filled lower band for spinless fermions
with a gap $\sim\Delta$ separating the lower and the upper bands. The Fermi
level, lying in the middle, is shifted when increasing $H$. $H_{c1}$ refers 
to the situation where the
Fermi energy crosses the gap and just lies at the bottom of the upper band. 
We have the relation $\Delta=g\mu_B H_{c1}$. Above $H_{c1}$, the upper band
is partially filled leading to an amount in the magnetization $\delta
m=m-1/2$. The problem is very similar to the 
commensurate-incommensurate transition\cite{D}.  
Near the plateau $\delta m\propto
\sqrt{\left|H-H_{c1}\right|}$ and
$1/{\cal T}_1\sim
T^{-1/2}$. The system remains {\it insulating} since  
$\Theta_-$ has a nonzero expectation value
and $\Phi_-$ has a gap\cite{Chitra}.

In conclusion, we have presented a new Kondo ladder system by
coupling the
so-called Takhtajan-Babujan S=1 chain to a half-filled 1D
electron gas and obtained low-energy properties using quantum field
theory techniques. Through the underscreened Kondo effect, this gives novel
insights into 1D spin liquids:
the optical conductivity reveals
only a pseudogap at $k=\pi$ despite of the insulating state, a coherent magnon
peak in the dynamical spin susceptibility coexists with an incoherent
spinon background, the magnetization yields a fractional-like 
quantization. Such a Kondo state still occurs in the 
vicinity of the $\beta=1$ 
point, but when $J_{\parallel}(1-\beta)=m_t$
the S=1 chain is in the Haldane phase\cite{Haldane2} and the electron
gas is decoupled. These results extend the semi-classical
picture of ref.\cite{Tsvelik} which is applicable only for
$J_{\parallel}a/v_F\rightarrow 0$
and should explain the coexistence
between a Kondo singlet state and strong antiferromagnetic fluctuations
in some heavy fermion compounds such as $UPt_3$. More generally, this problem
belongs to the class of Luttinger liquids in active environments, a topic
of importance not only to the Kondo chain physics, but also to
e.g. striped phases and carbon nanotubes.
K.L.H. thanks B. Coqblin, A. Honecker, T.M. Rice and A. Tsvelik for 
conversations.

\end{document}